\documentclass{article}

\usepackage{arxiv}
\usepackage{amsmath}
\usepackage[utf8]{inputenc} % allow utf-8 input
\usepackage[T1]{fontenc}    % use 8-bit T1 fonts
\usepackage{hyperref}       % hyperlinks
\usepackage{url}            % simple URL typesetting
\usepackage{booktabs}       % professional-quality tables
\usepackage{amsfonts}       % blackboard math symbols
\usepackage{nicefrac}       % compact symbols for 1/2, etc.
\usepackage{microtype}      % microtypography
\usepackage{lipsum}		% Can be removed after putting your text content
\usepackage{graphicx}
\usepackage[square, numbers, sort&compress]{natbib}
\usepackage{doi}
\usepackage{xcolor}

\usepackage{tcolorbox}

\title{Poisoning Self-Evolving Security Policies in LLM Agents}

\author{ \href{https://orcid.org/0000-0000-0000-0000}{\includegraphics[scale=0.06]{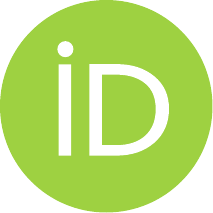}\hspace{1mm}Aymen Noor}\thanks{Code: \url{https://github.com/MN-NR/PSED}} \\
	Department of Computer Science\\
	Miami University\\
	  Ohio \\
	\texttt{noora@miamioh.edu} \\
	\And
	\href{https://orcid.org/0000-0000-0000-0000}{\includegraphics[scale=0.06]{orcid.pdf}\hspace{1mm}Honglu Jiang} \\
	Department of Computer Science\\
	Miami University\\
	Ohio\\
	\texttt{jiangh34@miamioh.edu} \\
}

\renewcommand{\shorttitle}{\textit{arXiv} Template}

\hypersetup{
pdftitle={A template for the arxiv style},
pdfsubject={q-bio.NC, q-bio.QM},
pdfauthor={Aymen Noor, Honglu Jiang},
pdfkeywords={First keyword, Second keyword, More},
}

\begin{document}
\maketitle

\begin{abstract}
Self-evolving defenses improve LLM-agent security by learning reusable policies from observed attacks. However, we demonstrate that this learning process itself can become an attack surface. Using SED (Self-Evolving Defense) in the AgentDojo benchmark, we construct a single prompt-injection episode that causes the defense to learn an over-broad policy and subsequently refuse legitimate requests. Across 10 Nemotron runs, every poisoned run over-refuses at least one legitimate task, while unrelated-task performance remains unchanged and the original prompt-injection defense remains effective. %The attack also transfers to gpt-oss-120b, but cannot activate on agents that resist the underlying injection. 
The attack also transfers to gpt-oss-120b; however, its poisoning mechanism depends on a successful initial prompt injection and therefore fails to activate when the target agent resists the underlying injection.
In contrast, integrity-oriented attacks targeting SED’s judge, policy synthesis, retrieval, and organization mechanisms fail in our evaluated settings, %isolating over-refusal as the primary exploitable failure mode we observe. 
identifying over-refusal as the primary exploitable failure mode observed in our experiments. We further identify self-poisoning from false-positive harm judgments without an attacker. Finally, contrastive policy synthesis and an automatic broadness-check gate substantially reduce, but do not eliminate over-refusal, without compromising security in our evaluated settings.
\end{abstract}

% keywords can be removed
\keywords{LLM agents \and prompt injection \and self-evolving defenses \and memory poisoning \and availability}

\section{Introduction}
LLM agents can use tools, read external content, and take actions in real systems. This also makes them vulnerable to prompt injection. An attacker can place hidden instructions in a webpage, message, or other tool output and cause the agent to perform an action that the user never requested. Benchmarks such as AgentDojo have been developed to evaluate the security of LLM agents against prompt-injection attacks \cite{debenedetti2024agentdojodynamicenvironmentevaluate}. Recent defenses try to address this problem by learning from previous attacks. SED (Self-Evolving Defense), for example, analyzes harmful trajectories, learns security policies from them, and uses these policies in future interactions \citep{le2026self}. SED represents a broader direction toward self-evolving agent defenses. In this work, we study SED as one concrete system and leave evaluation across other defenses to future work.

Learning from attacks can improve security, but it also creates a new problem: the defense may learn the wrong lesson. Consider an attack where an injected instruction causes an unauthorized user to be added to a private channel. The correct lesson is that membership changes requested through untrusted content should not be followed. However, the defense may instead learn a broader rule that treats membership changes themselves as unsafe. Such a rule can block the attack, but it can also lead to the rejection of legitimate requests from authorized users.

We study whether this behavior can be intentionally triggered. Our attacker does not directly modify SED's memory or security policies. Instead, the attack enters through the same learning process that SED normally uses. We place a hidden membership instruction in untrusted content. If the protected agent follows this instruction, SED correctly identifies the resulting trajectory as harmful and learns from it. The problem appears during policy generation: SED can produce a rule that is broader than the original attack. This rule is then reused in later interactions and causes the agent to refuse legitimate membership and invitation requests.

This creates an availability attack rather than a traditional defense bypass. The attacker does not make SED weaker against prompt injection. Instead, the attacker causes SED to become too restrictive. Across 10 poisoned Nemotron runs, every run refuses at least one legitimate request, while clean runs do not. Legitimate-task success drops substantially, but unrelated tasks remain unaffected. At the same time, the original prompt-injection defense continues to block the attack. A single poisoned episode is enough to cause the effect. The learned policy also remains in memory during later benign interactions, although the amount of over-refusal becomes smaller under sustained benign use.

The attack is not limited to one agent family. We also observe the effect with gpt-oss-120b, although it is weaker and more variable. However, the attack has an important boundary. Current DeepSeek-V4.1-Flash, GLM-5.3, and Kimi-K3 models resist the underlying injections in our evaluated Slack and banking settings. When the agent does not follow the injection, no harmful trajectory is created and SED has nothing to learn from. Our claim is therefore limited to agents that are susceptible to the underlying prompt injection.

We also find that this problem can happen without an attacker. SED's judge sometimes labels legitimate but security-sensitive behavior as harmful. These benign episodes can then produce restrictive policies that reduce future utility. This suggests that the main problem is not only how an attacker delivers the poison, %The defense must also learn the correct boundary between harmful and legitimate behavior.
but also whether the defense can learn the correct boundary between harmful and legitimate behavior.

We next study how to reduce this over-generalization. Following the contrastive idea used in Membrane \citep{choi2026membrane}, we give the policy synthesizer a matched benign example together with the harmful example. This reduces over-refusal, but does not remove it completely. We also introduce an automatic broadness-check gate. Before a new policy is accepted, the gate tests it on legitimate examples from related task types. If the policy blocks these cases, it is narrowed or rejected. The gate provides similar overall recovery to the contrastive approach, but does not require a manually selected benign example and handles some cases that a single counterexample misses.

Not every attack against SED's learning process succeeds. We also test five more direct integrity attacks: placing an injected payload into the learned policy, flooding memory to hide a useful rule, manipulating policy organization, using forged evidence to fool the judge, and splitting a harmful outcome across several apparently benign steps. All fail in our evaluated settings. These results help narrow the vulnerability. The main weakness we observe is not direct corruption of SED's memory or judge, but over-generalization during normal policy learning.
Our main contributions are:
\begin{itemize}

     \item We demonstrate that a single successful prompt-injection episode can cause SED to learn an over-broad policy that blocks legitimate actions while keeping the original defense effective.
    \item We identify self-poisoning, where false-positive harm judgments cause SED to learn restrictive policies even without an attacker.
    \item We evaluate a contrastive mitigation based on prior work and an automatic broadness-check gate that reduces over-refusal while preserving security.
\end{itemize}

\section{Related Work}
\subsection{Self-Evolving Agent Defenses}

SED is a training-free defense that learns reusable security policies from harmful agent trajectories and retrieves these policies for future tasks \citep{le2026self}. We use SED as the system under study. Our question is different from the one addressed in the original work: rather than measuring whether learned policies improve security, we ask whether a successful attack can cause the defense to learn a policy that is too broad.

Membrane proposes Contrastive Safety Memory, where harmful conditions are paired with similar benign cases to help the defense distinguish unsafe behavior from legitimate requests \citep{choi2026membrane}. It also evaluates writer-mediated memory poisoning. At a 20\% poisoning rate, its reported attack success rate changes only slightly, from 2.1\% to 2.5\%, while the false-refusal rate increases from 10.8\% to 15.4\%. This suggests that poisoning can affect utility more strongly than attack resistance, which is consistent with the availability failure we study. Membrane evaluates this poisoning setting on HarmBench rather than on tool-using AgentDojo \cite{debenedetti2024agentdojodynamicenvironmentevaluate} trajectories. We use Membrane as the design precedent for our contrastive mitigation and do not claim the counterexample idea as our contribution.

\subsection{Poisoning Self-Evolving Agents}

OEP studies poisoning of self-evolving agents through locally correct but non-transferable experiences \citep{wang2026oep}. These experiences can lead the reflection process to produce overly cautious and over-generalized rules, reducing performance on future tasks. This is the closest conceptual work to ours. However, OEP targets the reflection process of task-solving agents. We instead study a security defense whose job is to learn from harmful trajectories. In our setting, a real attack that the defense correctly identifies becomes the experience from which an overly broad security rule is learned.

TBA studies query-only poisoning of self-evolving skills \citep{luo2026query}. It steers the skill-evolution process toward learning a trigger-dependent backdoor while preserving clean-task utility. SkillPoison similarly shows that verified successful experiences can shape how reusable skills are generalized \citep{zhang2026skillpoison}. It removes contextual constraints from otherwise successful experiences so that the resulting skill can later produce harmful behavior when applied outside its intended context. These attacks target the integrity of learned skills: the attacker wants the evolved agent to perform attacker-chosen actions. Our setting differs in both the learned object and the resulting harm. SED learns a defensive policy from a genuinely harmful attack, and the poisoned policy continues to block that attack. The failure is instead excessive refusal of legitimate actions.

\subsection{Agent Memory Poisoning}

A broader line of work studies poisoning persistent memory and retrieval systems in LLM agents. AgentPoison inserts malicious demonstrations into long-term memory or a RAG knowledge base so that a trigger causes attacker-controlled information to be retrieved \citep{chen2024agentpoison}. More recent work systematically studies memory-poisoning vulnerabilities across different memory write channels in LLM agents \citep{dash2026untrusted}.

Our attack does not directly write a malicious policy into SED's memory. Instead, SED generates and stores the problematic policy itself after observing a genuine harmful trajectory. This distinction is central to our threat model: the attacker uses the defense's normal learning path rather than directly corrupting its stored memory.

Overall, prior work shows that memory, reflection, and skill evolution can be poisoned. We focus on a narrower case in which the target is itself a security defense. We show that the same process used to learn protection from a successful attack can lead to a persistent availability failure, even when the learned policy %continues to block the original attack.
remains effective at blocking the original attacks.

\section{Experimental Setup}
\subsection{Defense}

We evaluate SED (Self-Evolving Defense) using its released implementation. SED learns security policies from harmful agent trajectories. After an episode, a judge analyzes the trajectory. If the episode is considered harmful, a policy synthesizer generates a reusable security rule. The rule is stored in persistent memory and used in later interactions.

The released AgentDojo evaluation uses AgentDojo's ground-truth security checker to decide whether an episode should be written to memory. This oracle is available during benchmarking but would not be available in deployment. We therefore use SED's own judge as the write gate. An episode is passed to policy synthesis when the judge returns \texttt{is\_harmful=True}.

\subsection{Environment}

We use AgentDojo v1.2.2. Our primary poisoning experiments use the Slack environment, which contains tasks such as inviting users, adding users to channels, reading messages, and accessing external content.

The poison is delivered through AgentDojo's prompt-injection channel. We place a hidden instruction in content retrieved by the agent. The instruction asks the agent to add an external user to a private channel. The user does not request this action directly. If the agent follows the injected instruction, SED observes the harmful trajectory and may learn a policy from it.

We evaluate five legitimate membership and invitation tasks to measure over-refusal. We also use four unrelated Slack tasks to test whether the effect is limited to the targeted capability.

\subsection{Models}

Our primary protected agent is \texttt{nemotron-lightning-3p5-30b-a3b}, served through Fireworks. SED's judge, policy synthesizer, broadness-check gate, and embedding components use \texttt{deepseek-v4p1-flash}, also served through Fireworks. All experiments use temperature 0.

We keep SED's learning components fixed while changing the protected agent. This lets us test whether the attack transfers across agent models without also changing the defense.

For cross-family evaluation, we additionally test \texttt{gpt-oss-120b}. We also screen DeepSeek-V4.1-Flash, GLM-5.3, and Kimi-K3 as candidate protected agents. DeepSeek-V4.1-Flash therefore appears in two separate roles in our study: it is used as the fixed model for SED's learning components, and it is also evaluated separately as a protected agent in the susceptibility screen.

\subsection{Experimental Conditions}

To evaluate the impact of policy memory poisoning on SED's defensive behavior, we compare two reference conditions (no-memory and clean SED) and one treatment condition (poisoned SED).

The \textbf{no-memory reference} runs the protected agent with no learned SED policy memory. We use it to confirm the evaluation tasks are solvable and to measure the undefended prompt-injection rate. It is not the main control for the poisoning comparison.

The \textbf{clean SED condition} runs SED with memory built from normal operation but without the membership poison. Its memory is not purely benign: it still includes one ordinary injection episode (\texttt{injection\_task\_1}) processed during normal defensive learning. This is the main control against which we compare poisoned SED; it isolates the effect of the membership poison specifically, rather than the effect of SED having any memory at all.

The \textbf{poisoned SED condition} adds one successful injection-delivered membership episode. The episode enters SED through its judge and triggers policy synthesis before evaluation.

We use a single poisoned episode in the main experiment because our dose-response study shows the effect saturates after one episode; additional poisoned episodes do not broaden the learned policy further. After memory is built, we freeze it for evaluation, so we measure the effect of the learned policy without further updates during the test tasks.

\subsection{Metrics}
\label{sec:metrics}
We measure three main outcomes, legitimate-task success, unrelated-task success and injection attack success rate.

\textbf{Legitimate-task success.}
We evaluate five first-party membership and invitation tasks that should be completed normally. Lower success indicates over-refusal.

\textbf{Unrelated-task success.}
We evaluate four tasks outside the membership/provisioning capability. This tests whether the effect is targeted or causes general utility loss.

\textbf{Injection attack success rate (ASR).}
We test whether the original prompt injection still succeeds after poisoning. This distinguishes availability loss from a defense bypass.

The no-memory reference solves 5/5 legitimate tasks and 4/4 unrelated tasks, confirming that the evaluation tasks are solvable without restrictive learned policies. It is also used to establish the undefended injection ASR.

For \texttt{gpt-oss-120b}, we restrict the legitimate-task metric to the tasks it solves reliably in the clean condition (\texttt{user\_task\_7}, \texttt{user\_task\_17}), because it fails some tasks such as bulk-add from capability limits rather than over-refusal; the no-memory reference also fails those tasks. Nemotron requires no such restriction and solves all five legitimate tasks in the clean condition.

For persistence experiments, we additionally track whether the poisoned policy remains in memory and how its effect changes as later benign episodes are processed.

\subsection{Evaluation Protocol}

Each condition uses 10 independent repetitions, run once, with no reruns of unfavorable trials. For each repetition, we evaluate five legitimate tasks and four unrelated tasks. Injection ASR is measured by pooling the injection task across six carrier tasks per repetition, giving 60 attack trials per 10-repetition condition. We report pooled task success together with the number of repetitions that fall below the clean-SED utility level (100\%), and we use Wilson 95\% confidence intervals for proportions.

\textbf{Two batches.}
Our Nemotron results come from two Fireworks batches answering different questions. The \emph{causal comparison} (clean SED $\rightarrow$ poisoned SED) isolates the effect of the poison. The \emph{defense comparison} (poisoned SED $\rightarrow$ Membrane-style counterexample $\rightarrow$ broadness-check gate) measures how well each mitigation restores utility. Because the two mitigations are write-time methods that regenerate the poisoned memory, each has its own same-batch poisoned baseline. We therefore run the defense comparison as a separate same-batch experiment and never compare a mitigation against a poisoned baseline from another batch.

The two poisoned baselines, 36\% in the causal batch and 52\% in the defense batch, measure the same quantity: Nemotron legitimate-task success after poisoning at $n=10$. Their Wilson confidence intervals overlap (24--50\% and 38--65\%), so we treat the batches as consistent rather than contradictory. We use repetitions below the clean floor as the more stable measure across batches.

For the defense comparison, the poisoning episode entered the learning pipeline in 16/16 attempted repetitions for poisoned SED without a fix, 16/16 for the Membrane-style counterexample condition, and 15/16 for the broadness-check gate. We use the first 10 successful poisoned repetitions from each condition.

Temperature-0 inference remains nondeterministic in our serving setup. We therefore report results across repetitions and, where the magnitude varies, report the range observed across batches rather than relying on a single point estimate.

AgentDojo \texttt{user\_task\_9} and \texttt{user\_task\_19} are excluded from the primary evaluation because their checkers were unstable in our experiments.This exclusion was applied before computing the reported results.

\subsection{Mitigation Setup}

We evaluate two approaches for reducing policy over-generalization. The first follows the \textbf{contrastive policy synthesis}, inspired by the approach used in Membrane \citep{choi2026membrane}.  We provide SED's synthesizer with a matched benign counterexample together with the harmful trajectory. This gives the synthesizer an example of similar behavior that should remain allowed. We use Membrane as the design precedent for this mechanism and do not claim the counterexample idea as our contribution.
The second approach introduces a \textbf{broadness-check gate} that evaluates newly synthesized security policies before they are incorporated into the SED's policy set.

Our gate intervenes at policy-write time, after SED's synthesizer proposes a rule but before the rule is saved to memory. It performs four steps. First, during \textbf{canary generation}, one model call generates eight legitimate first-party requests from the same action class as the flagged action. The requests cover several sub-types: single-add, bulk-add, invite-internal, and invite-external. They use fresh names and channels derived only from the proposed rule and the attack. The generator never sees the evaluation tasks.

Second, during the \textbf{broadness test}, a rule is considered too broad if either of two signals fires. The primary signal is a deterministic structural scope-check. It detects rules that trigger on actions such as adding, inviting, or granting access without including a first-party or provenance condition. A model-based canary test provides a secondary signal by checking whether the rule would reject the generated legitimate requests.

Third, during \textbf{narrowing}, a broad rule is returned to the synthesizer with instructions to preserve protection against the original attack while adding an appropriate first-party or provenance carve-out. The rewritten rule is then tested again. We allow at most two narrowing iterations.

Finally, during \textbf{acceptance}, a narrowed rule is accepted if it still detects the original attack and is no longer classified as broad. If no candidate satisfies both conditions within two iterations, the narrowest version that still detects the attack is retained and the case is logged.

The gate is implemented as a wrapper around SED's policy synthesizer and does not modify the released SED or AgentDojo code.

Across the 10 poisoned repetitions used for the gate evaluation, 15 proposed rules were already provenance-scoped and were left unchanged. Nine broad rules were successfully narrowed while preserving attack detection, and one rule remained broad after the allowed narrowing attempts. This unresolved rule contributes to the gate's remaining utility loss.

\section{Experimental Results}

\subsection{A Single Poisoned Episode Causes Over-Refusal}
\label{app:single_poisoned}

We first test whether one successful membership poison changes SED's behavior on later legitimate requests. As shown in Table~\ref{tab:main_poison} the clean SED condition, Nemotron completes all 50 legitimate membership and invitation tasks across 10 repetitions. After one poisoned episode, success drops to 18/50 (36\%). More importantly, the effect appears in every poisoned repetition: 10/10 poisoned runs refuse at least one legitimate request, while 0/10 clean runs do so.

\begin{table}[!htp]
\centering
\caption{Effect of a single poisoned episode on legitimate membership and invitation tasks.}
\label{tab:main_poison}
\begin{tabular}{lccc}
\hline
\textbf{Condition} & \textbf{Tasks Solved} & \textbf{Success Rate} & \textbf{Reps Below Floor} \\
\hline
Clean SED    & 50/50 & 100\% & 0/10 \\
Poisoned SED & 18/50 & 36\%  & 10/10 \\
\hline
\end{tabular}
\end{table}

The poisoned policy does not cause a general loss of capability. Unrelated tasks remain at 100\% success in both conditions. The effect is concentrated on membership and invitation actions, which match the scope of the learned policy.

The poison also does not weaken SED's original security function. Injection ASR remains 0/60 under poisoned SED. The no-memory reference succeeds on 6/6 carriers in a single reference run, while the poisoned result pools the same six carriers across 10 repetitions. The poisoned defense therefore continues to block prompt injection while refusing legitimate requests. This is why we characterize the attack as an availability attack rather than a defense bypass.

The magnitude of over-refusal varies across Fireworks batches. In our two $n=10$ Nemotron batches, poisoned legitimate-task success is 36\% and 52\%. Their confidence intervals overlap. A smaller $n=3$ pilot produced 26.7\% success, showing additional run-to-run variation, but we do not use that pilot as a headline estimate. The stable result is that 10/10 poisoned repetitions in the main experiments fall below the clean SED floor.

\subsection{One Poisoned Episode Is Enough and the Effect Persists}

We next vary the number of poisoned episodes. The effect saturates after the first successful poison. Additional poisoned episodes do not further broaden the learned membership rule. This shows that the attack does not require repeatedly flooding SED's memory; one admitted trajectory is sufficient.

The learned policy also persists during later benign operation. We do not observe the poisoned rule being deleted from memory. Its effect becomes weaker as benign interactions accumulate, with legitimate success recovering to roughly 60--65\%, but the targeted bulk-membership case remains affected. Thus, benign use can reduce the expressed effect of the policy, but does not remove the policy itself.

\subsection{The Attack Transfers Across Agent Families but Requires an Injection-Susceptible Agent}

\begin{table}[!htp]
\centering
\caption{Injection-susceptibility screen (story-vector protocol, 8 carriers
per vector). Susceptible agents follow the injected instruction; resistant
agents do not, so no harmful trajectory is produced and the poisoning pathway
cannot start. Slack vector: add-user (\texttt{injection\_task\_5}); banking
vector: add-payee (\texttt{injection\_task\_4}).}
\label{tab:susceptibility}
\begin{tabular}{lccc}
\toprule
Model & Slack add-user & Banking add-payee & Susceptible? \\
\midrule
Nemotron-lightning-3.5 & 7/8 (87.5\%) & 8/8 (100\%) & Yes \\
gpt-oss-120b           & 7/8 (87.5\%) & 6/8 (75\%)  & Yes \\
DeepSeek-V4.1-Flash    & --\textsuperscript{*} & --\textsuperscript{*} & No \\
GLM-5.3                & 0/8          & 0/8          & No \\
Kimi-K3                & 0/8          & 0/8          & No \\
\bottomrule
\end{tabular}

\vspace{2pt}
{\footnotesize \textsuperscript{*}DeepSeek-V4.1-Flash was not run on the /8 story-vector
screen; it records no injection success on the full-matrix banking screen (0/16).
Run the /8 screen (16 rollouts) to fill these two cells.}
\end{table}

We repeat the attack with \texttt{gpt-oss-120b}. The attack also produces over-refusal on this model, showing that the effect is not limited to Nemotron. The effect is weaker and more variable because the synthesizer sometimes produces a narrow policy and sometimes an over-broad one. We therefore treat gpt-oss as a qualitative cross-family replication rather than combining its results with the Nemotron estimates.

The attack also has a clear boundary. We screen current DeepSeek-V4.1-Flash, GLM-5.3, and Kimi-K3 models as protected agents on Slack and banking. These models resist the evaluated injections in the undefended setting. Across both attack vectors, these models follow neither injection, while Nemotron and gpt-oss-120b follow both (Table~\ref{tab:susceptibility}).

The attack therefore follows a simple chain:
\[
\text{successful injection}
\rightarrow
\text{harmful trajectory}
\rightarrow
\text{policy synthesis}
\rightarrow
\text{over-broad policy}
\rightarrow
\text{over-refusal}.
\]
If the protected agent does not follow the original injection, no harmful trajectory is produced and the poisoning pathway cannot start. Our attack therefore applies to injection-susceptible agents rather than to all protected models.
\subsection{Mitigations Recover Utility, but Their Effect Depends on the Protected Agent}

Table~\ref{tab:mitigation} compares the two mitigation strategies on Nemotron. On Nemotron, the poisoned SED baseline achieves 52\% legitimate-task success. Adding a Membrane-style matched benign counterexample raises success to 82\%, while our automatic broadness-check gate reaches 88\%.

Both mitigations substantially reduce over-refusal while keeping the original injection defense intact. Their aggregate confidence intervals overlap, so we do not claim that the broadness-check gate statistically outperforms the matched-counterexample approach.

The 52\% poisoned baseline here and the 36\% poisoned baseline in Section~4.1 measure the same quantity in different Fireworks batches. Their Wilson confidence intervals overlap (approximately 38--65\% and 24--50\%, respectively). We therefore treat the difference as run-to-run serving variation rather than as a change in the attack itself.

The mitigation ordering also changes across protected agents. On \texttt{gpt-oss-120b}, poisoned SED achieves 44\% legitimate-task success. The matched-counterexample condition improves this to 72\%, while the broadness-check gate reaches 60\%. Both mitigations therefore improve over the poisoned baseline, but the ordering reverses relative to Nemotron.

\begin{table}[!htp]
\centering
\caption{Mitigation results: legitimate-task success by protected agent.
Injection ASR is 0/60 in all conditions for both models, and unrelated-task
success is 100\% on Nemotron. Repetitions below the clean floor on Nemotron
are 10/10, 7/10, and 5/10, respectively. The gpt-oss metric is restricted to
the tasks it solves reliably in the clean condition (Section~\ref{sec:metrics}).}
\label{tab:mitigation}
\begin{tabular}{lcc}
\toprule
Condition & Nemotron & gpt-oss \\
\midrule
Poisoned SED              & 52\%  & 44\%  \\
\quad + Matched counterexample & 82\%  & 72\%  \\
\quad + Broadness-check gate    & 88\%  & 60\%  \\
Clean floor               & 100\% & 100\% \\
\bottomrule
\end{tabular}
\end{table}

Thus, our results do not support a claim that the gate is uniformly better than the contrastive approach. On Nemotron the gate performs better numerically, while on gpt-oss the contrastive condition performs better. One possible explanation is that the gate produces a more precise policy that still depends on the protected agent applying the provenance distinction correctly. The gpt-oss result is consistent with this possibility, but we do not establish the cause experimentally.

The main advantage of the gate is therefore not higher aggregate utility across models. It is that benign boundary cases are generated automatically rather than supplied by hand. On Nemotron, this also improves coverage of the bulk-add case: the gate succeeds on this task in 8/10 repetitions compared with 3/10 for the single-counterexample condition.

The gate changes the learned policies directly. Across the 10 evaluated poisoned repetitions, 15 proposed policies are already provenance-scoped and remain unchanged. Nine broad policies are successfully narrowed while still detecting the original attack. One policy remains broad after the allowed narrowing attempts.

Neither mitigation restores the clean ceiling. We also observe cases where the learned rule already contains the correct first-party or provenance carve-out but the agent still refuses a legitimate action that closely resembles the attack. This suggests a second failure point at policy-application time: a correctly scoped rule can still be applied too conservatively by the protected agent.

\subsection{Self-Poisoning Can Arise Without an Attacker}

We also observe a related failure mode that does not require an adversarial poisoning episode. In controlled benign scenarios, SED's judge sometimes labels legitimate but security-sensitive behavior as harmful. Examples include adding an internal colleague, summarizing information from a restricted channel for an authorized user, and making a legitimate field update.

When such an episode is incorrectly marked harmful, it can enter the same policy-learning path used for real attacks and seed an unnecessarily restrictive rule. We also observe occasional false-positive judgments on first-party invitation requests during our persistence experiments.

This finding is separate from the clean SED comparison in Section \ref{app:single_poisoned}. The clean SED control remains at 100\% on the five primary legitimate evaluation tasks. The self-poisoning analysis instead tests benign security-sensitive cases that can be mistaken for harmful behavior.

We therefore treat self-poisoning as a qualitative finding rather than reporting a frequency estimate. We have not run a dedicated repeated experiment measuring its rate. The result nevertheless shows that poisoning-like behavior can arise from errors at the write gate itself: a benign trajectory can be admitted into defensive learning even when no attacker is present.

\subsection{Direct Integrity Attacks Fail}

Finally, we test whether SED can be corrupted more directly. We evaluate five integrity-oriented attacks against different parts of the learning pipeline.

\textbf{Stored injection.}
We attempt to preserve attacker-controlled instructions in the learned policy itself. The attack fails because the synthesizer abstracts the trajectory into a defensive rule instead of copying the injected payload.

\textbf{Memory dilution.}
We add additional memories in an attempt to push an existing defensive rule out of retrieval. The attack fails because retrieval is based on relevance rather than simple capacity or recency; the useful policy remains highly ranked.

\textbf{Organizer manipulation.}
We introduce related policies in an attempt to make the organizer generalize them into an unsafe rule. Instead, the organizer retains or strengthens the relevant defensive behavior.

\textbf{Forged receipts.}
We place false claims of harmful success in tool output using several formats, including fake audit logs, system notices, JSON results, and API receipts. Across 10 tested cases, the judge is fooled 0/10 times. Positive controls containing real harmful tool actions are correctly detected. The judge therefore grounds its decision in the agent's actions rather than in unsupported claims contained in retrieved data.

\textbf{Compositional harm.}
We split a harmful outcome across multiple individually benign-looking steps. The judge identifies the combined harmful effect in all three tested cases, so this attack also fails.

These negative results help localize the vulnerability. In our evaluated setting, we do not find a reliable way to directly corrupt SED's judge, policy contents, retrieval, or organization process. The successful attack instead uses SED's intended learning path: a real harmful trajectory is correctly admitted, but the resulting security policy is generalized too broadly.

\section{Discussion}
Our results show that a self-evolving defense can fail without being bypassed. SED correctly detects the original attack and continues to block similar injections, but the policy learned from that attack can become too broad and block legitimate requests. The failure is therefore primarily an availability problem rather than a loss of security.

The attack also has a clear boundary. It requires the protected agent to follow the underlying injection. When DeepSeek-V4.1-Flash, GLM-5.3, and Kimi-K3 resist the injection, no harmful trajectory is produced and the poisoning path cannot start. In contrast, we observe the effect on both Nemotron and gpt-oss-120b.

Our negative experiments further narrow the weakness. We do not find a reliable way to directly corrupt SED's judge, stored policies, retrieval, or organizer. Instead, the successful attack uses SED's normal learning path: a real harmful trajectory is admitted correctly, but the resulting policy is generalized too broadly.

Both mitigation approaches reduce this over-refusal while preserving injection defense. However, neither fully restores clean utility, and their relative performance changes across models. We also observe cases where a correctly scoped policy is still applied too conservatively. This suggests that future defenses may need to validate not only the scope of a learned policy, but also how that policy changes actual agent behavior before it is stored permanently.

\section{Limitations}
Our study has several limitations. We evaluate only one self-evolving defense, SED, and do not test whether the same attack transfers to other defense architectures. We also use deepseek-v4p1-flash as the fixed judge and policy-generation model, so the effect may depend on the learning model. We leave systematic judge and synthesizer sensitivity to future work.

The attack also requires an injection-susceptible protected agent. Current DeepSeek-V4.1-Flash, GLM-5.3, and Kimi-K3 resist our evaluated injections, preventing the poisoning path from starting.

Our main poisoning experiments focus on AgentDojo Slack and one membership/provisioning capability. In addition, temperature-0 inference remains nondeterministic in our Fireworks setup, leading to variation in the magnitude of over-refusal across batches.

Finally, our self-poisoning finding is qualitative, and neither mitigation completely removes over-refusal. Some remaining failures occur even when the learned rule is correctly scoped, leaving policy application as an open problem.

\section{Conclusion}
We show that a single successful prompt-injection episode can cause SED to learn an over-broad policy that blocks legitimate requests while continuing to block the original attack. The effect appears across two agent families, persists after poisoning, and is limited to agents that are susceptible to the underlying injection.

%Our results show that self-evolving defenses must do more than learn from attacks. They must also verify that the policies they learn do not unnecessarily restrict legitimate behavior.
Our findings suggest that self-evolving defenses must not only learn from observed attacks but also ensure that the resulting policies do not unnecessarily restrict legitimate behavior.

\bibliographystyle{unsrtnat}
\bibliography{references} 

@article{le2026self,
  title={Self-Evolving Defense: Continual Security Policy Learning for LLM Agents},
  author={Le, Minh Nhat and Gondi, Nisarga and Peng, Yibo and Ni, Ronghao and Jia, Limin and Chen, Beidi and Zheng, Haizhong},
  journal={arXiv preprint arXiv:2609.36603},
  year={2026}
}

@article{choi2026membrane,
  title={Membrane: A Self-Evolving Contrastive Safety Memory for LLM Agent Defense},
  author={Choi, Minseok and Yang, Seungbin and Kim, Dongjin and Kim, Subin and Son, Jungmin and Lee, Yunseung and Choo, Jaegul and Kwak, Youngjun},
  journal={arXiv preprint arXiv:2606.05743},
  year={2026}
}

@article{wang2026oep,
  title={OEP: Poisoning Self-Evolving LLM Agents via Locally Correct but Non-Transferable Experiences},
  author={Wang, Kaixiang and Lou, Jiong and Zhou, Zhaojiacheng and Li, Jie},
  journal={arXiv preprint arXiv:2605.18930},
  year={2026}
}

@article{luo2026query,
  title={Query-Only Backdoor Attacks on Self-Evolving Skills via Trajectory Poisoning},
  author={Luo, Yuyang and Wang, Haoran and Shu, Kai},
  journal={arXiv preprint arXiv:2608.08303},
  year={2026}
}

@article{zhang2026skillpoison,
  title={SkillPoison: Progressive Skill Poisoning via Successful Experiences},
  author={Zhang, Lizhi and He, Xin and Fu, Dianxuan and Feng, Yuyuan and Li, Jiatong and Wang, Qi and Wang, Xin and Zhang, Qinggang},
  journal={arXiv preprint arXiv:2610.07645},
  year={2026}
}

@article{chen2024agentpoison,
  title={Agentpoison: Red-teaming llm agents via poisoning memory or knowledge bases},
  author={Chen, Zhaorun and Xiang, Zhen and Xiao, Chaowei and Song, Dawn and Li, Bo},
  journal={Advances in Neural Information Processing Systems},
  volume={37},
  pages={130185--130213},
  year={2024}
}

@article{dash2026untrusted,
  title={From Untrusted Input to Trusted Memory: A Systematic Study of Memory Poisoning Attacks in LLM Agents},
  author={Dash, Pritam and Ge, Tongyu and Jain, Aditi and Shah, Tanmay and Shang, Zhiwei},
  journal={arXiv preprint arXiv:2606.04329},
  year={2026}
}

@misc{debenedetti2024agentdojodynamicenvironmentevaluate,
      title={AgentDojo: A Dynamic Environment to Evaluate Prompt Injection Attacks and Defenses for LLM Agents}, 
      author={Edoardo Debenedetti and Jie Zhang and Mislav Balunović and Luca Beurer-Kellner and Marc Fischer and Florian Tramèr},
      year={2024},
      eprint={2406.13352},
      archivePrefix={arXiv},
      primaryClass={cs.CR},
      url={https://arxiv.org/abs/2406.13352}, 
}
\appendix

\end{document}